\documentclass[preprint,superscriptaddress]{revtex4-1}
\pdfoutput=1 
\usepackage{txfonts}
\usepackage{CJK} 
\usepackage{float}
\usepackage{here}
\usepackage{mathrsfs}
\usepackage{here}
\usepackage{mathrsfs}
\usepackage[dvipdfmx]{graphicx}
\usepackage{color}

\begin{document}
\begin{CJK*}{}{} 
\title{Optical determination of the exchange stiffness constant in an iron garnet}
\author{Keita~Matsumoto}
\affiliation{Department of Physics, Kyushu University, Fukuoka 819-0385, Japan}
\author{Thomas Br\"acher}
\affiliation{Fachbereich Physik and Landesforschungszentrum OPTIMAS, Technische Universit\"at Kaiserslautern, D-67663 Kaiserslautern, Germany}
\author{Philipp Pirro}
\affiliation{Fachbereich Physik and Landesforschungszentrum OPTIMAS, Technische Universit\"at Kaiserslautern, D-67663 Kaiserslautern, Germany}
  \author{Dmytro Bozhko}
  \affiliation{Fachbereich Physik and Landesforschungszentrum OPTIMAS, Technische Universit\"at Kaiserslautern, D-67663 Kaiserslautern, Germany} 
\author{Tobias Fischer}
\affiliation{Fachbereich Physik and Landesforschungszentrum OPTIMAS, Technische Universit\"at Kaiserslautern, D-67663 Kaiserslautern, Germany}
\affiliation{Graduate School Materials Science in Mainz, Gottlieb-Daimler-Strasse 47, D-67663 Kaiserslautern, Germany}
 \author{Moritz Geilen}
  \affiliation{Fachbereich Physik and Landesforschungszentrum OPTIMAS, Technische Universit\"at Kaiserslautern, D-67663 Kaiserslautern, Germany}
  \author{Frank Heussner}
  \affiliation{Fachbereich Physik and Landesforschungszentrum OPTIMAS, Technische Universit\"at Kaiserslautern, D-67663 Kaiserslautern, Germany}
\author{Thomas Meyer}
   \affiliation{Fachbereich Physik and Landesforschungszentrum OPTIMAS, Technische Universit\"at Kaiserslautern, D-67663 Kaiserslautern, Germany}
\author{Burkard Hillebrands}
\affiliation{Fachbereich Physik and Landesforschungszentrum OPTIMAS, Technische Universit\"at Kaiserslautern, D-67663 Kaiserslautern, Germany}
\author{Takuya~Satoh}
\affiliation{Department of Physics, Kyushu University, Fukuoka 819-0385, Japan}
\date{\today}
\begin{abstract}
	Brillouin light scattering measurements were performed in the backscattering geometry on a Bi-substituted rare earth iron garnet. We observed two different peaks, one attributed to a surface spin wave in the dipole-exchange regime. The other is referred to as a backscattering magnon mode, because the incident light in this case is scattered backward by exchange-dominated spin wave inside the material. We propose a method to estimate the exchange stiffness constant from the frequency of the backscattering magnon mode. The obtained value is comparable with the previously reported values for Y$ _3 $Fe$ _5 $O$ _{12} $.
\end{abstract}
	\maketitle
\end{CJK*}
	The exchange stiffness describes the strength of short-range coupling between two spins inside a magnetic material\cite{Stancil,SpinwaveConfinement}. It can control domain wall patterning\cite{Bobeck1969,McMichael1997}, which is used in magnetic storage devices\cite{Chappert2007}.
	Moreover, the exchange stiffness affects the frequency of spin waves\cite{Kittel1958,Kalinikos1986,Kalinikos1990}. Thus, the determination of the exchange stiffness plays an important role in applications such as magnetic random access memories and information transmitting media.
	
	In general, this quantification is not simple when compared to that of saturation magnetization. In previous studies, it was achieved by the analysis of perpendicularly standing spin wave (PSSW) modes\cite{Vernon84,Liu96,Raasch94,Hamrle09,Gaier2009,Eyrich12,Klingler15,Sebastian15}, neutron scattering experiments\cite{Shirane68,Passell1976}, X-ray magnetic linear dichroism spectroscopy\cite{Scholl04}, and analysis of $M$-$H$ curves\cite{Girt11}.
	The PSSW mode is a standing spin wave along the thickness. The standing wave in a thin film is known to have a large wavenumber; in this regime, the exchange interaction is dominant. Many reports estimate the exchange stiffness through the analysis of PSSWs.
	
	Brillouin light scattering (BLS) spectroscopy\cite{Demokritov2001} is a crucial method to detect spin waves including the PSSW modes. BLS results from the inelastic scattering of photons from spin waves under momentum and energy conservation; the scattered light carries frequency and wavenumber information of the spin waves that it interacted with.
	
	In recent times, a backscattering magnon (BSM) mode has been found in a BLS spectrum of the ferrimagnetic insulator yttrium iron garnet (YIG) at room temperature\cite{Agrawal13}. A spin wave with a wavenumber of $ 2nk_{\text{I}} $ collides with the incident photons, which are part of the initial incident light refracted by the material, possessing a wavenumber $nk_{\text{I}}$, where $ n$ is the refractive index and $ k_{\text{I}}$ is the wavenumber of the incident photon in vacuum. The scattered photons then propagate back along their original path since they have the wavenumber of $-nk_{\text{I}}\ (=nk_{\text{I}}-2nk_{\text{I}}) $. In addition, $ k_{\text{I}}$ is generally much larger than the wavenumber of the spin wave under the magnetostatic approximation (dipolar-dominated spin wave). Thus, the BSM mode is the exchange-dominated wave with the maximum wavenumber that can be detected by a BLS experiment\cite{Agrawal13}. Unlike PSSWs, its resonance frequencies are not influenced by the film thickness or pinning conditions.
	
	In this study, we propose a scheme to determine the exchange stiffness constant $ A_{\text{ex}} $ by studying the BSM mode using BLS spectroscopy. We investigated the dependence of the BLS peaks on the external magnetic field strength and the angle of incidence of the probing light. The observed peaks were analyzed using a model of (dipole-)exchange spin waves so that the observed spin wave mode could be assigned. This fitting enables us to determine the exchange stiffness constant $A_{\text{ex}}$ accurately. It is found to be comparable with that of YIG that has the same crystalline structure as our sample.
	
\begin{figure}[htbp]
		\includegraphics[width=10cm]{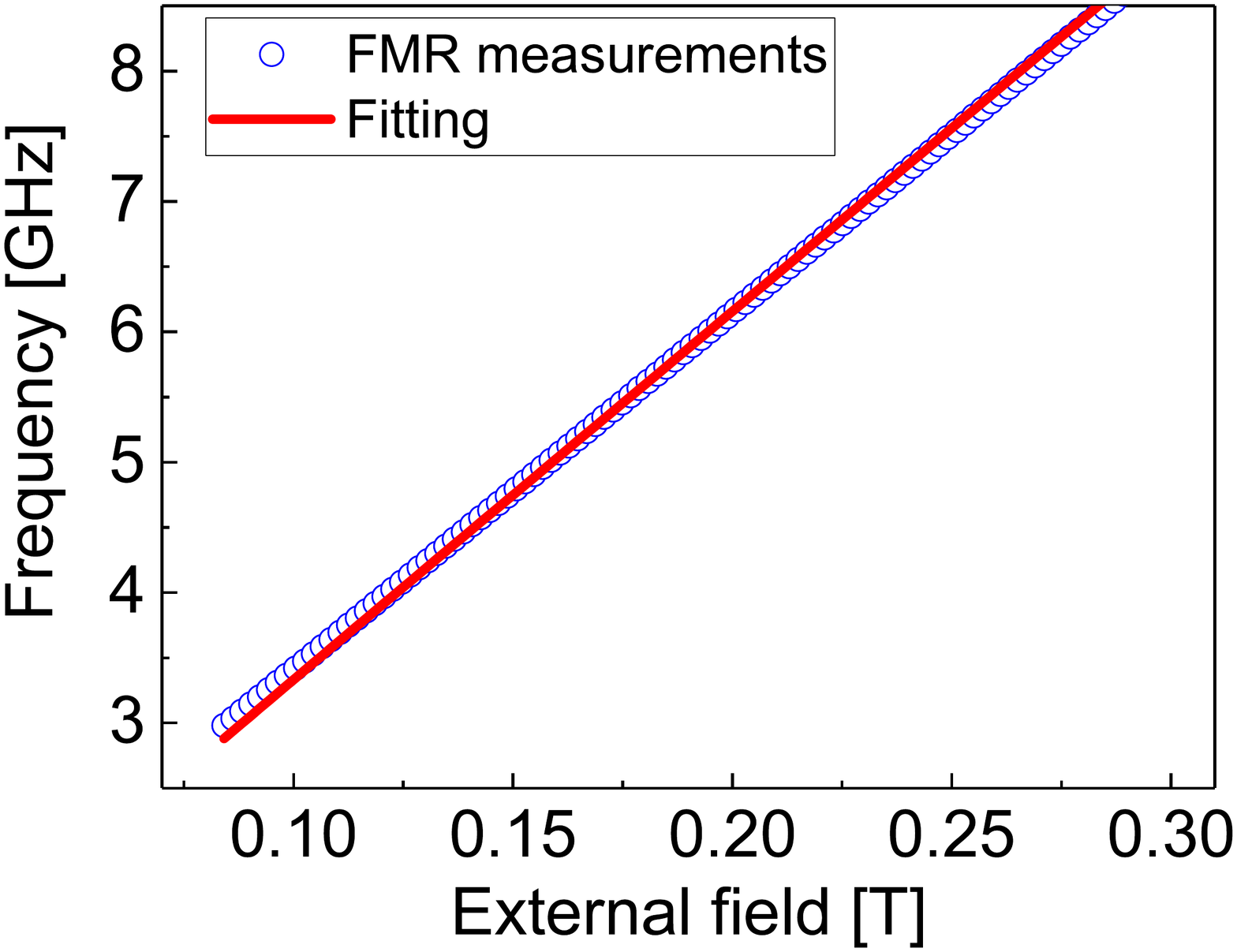}
		\caption{(Color online) The in-plane field dependence of the FMR spectrum (blue dots). The fitting result (dashed line).}
		\label{FMR}
	\end{figure}

	Our sample was a (111)-oriented 140 $\mu$m-thick Bi-doped rare earth iron garnet film \\(Gd$ _{3/2}$Yb$ _{1/2}$BiFe$_5$O$_{12}$), grown on a gadolinium gallium garnet substrate by the liquid-phase epitaxy method. The sample was a ferrimagnetic single crystal with a Curie temperature of 573 K. This compound is known to show a large Faraday rotation and highly efficient spin wave excitation owing to the inverse Faraday effect\cite{Satoh12,Parchenko13,Yoshimine14,Yoshimine17,Chekhov18}, which means it exhibits strong magneto-optical coupling. Thus, we can expect a pronounced magnetic BLS signal.
	The material parameters are -- saturation magnetization $M_{\text{s}}=0.114$ T$/\mu_0$\cite{Satoh12}, gyromagnetic ratio $ \gamma=28 $ GHz/T, and sample thickness $d=140$ $\mu$m with lateral size of 5.5 mm $\times$ 5.5 mm.
	In order to measure the perpendicular anisotropy field $ H_{\text{u}}$, ferrimagnetic resonance (FMR) measurement was performed and the data are shown in Fig. \ref{FMR}. Fitting with Kittel's formula\cite{Lenk11}$ f=\gamma\mu_0\sqrt{H_{\text{ext}}(H_{\text{ext}}+M_{\text{s}}-H_{\text{u}})}$ yields $\mu_0(M_{\text{s}}-H_{\text{u}})=0.042$ T, where $H_{\text{ext}}$ is an in-plane magnetic field.
	In addition to these parameters, we used the value of refractive index as $n=2.8$ at 532 nm, which is the value measured independently by Doormann {\it et al}.\cite{Doormann84}
	
\begin{figure}[htbp]
		\includegraphics[width=10cm]{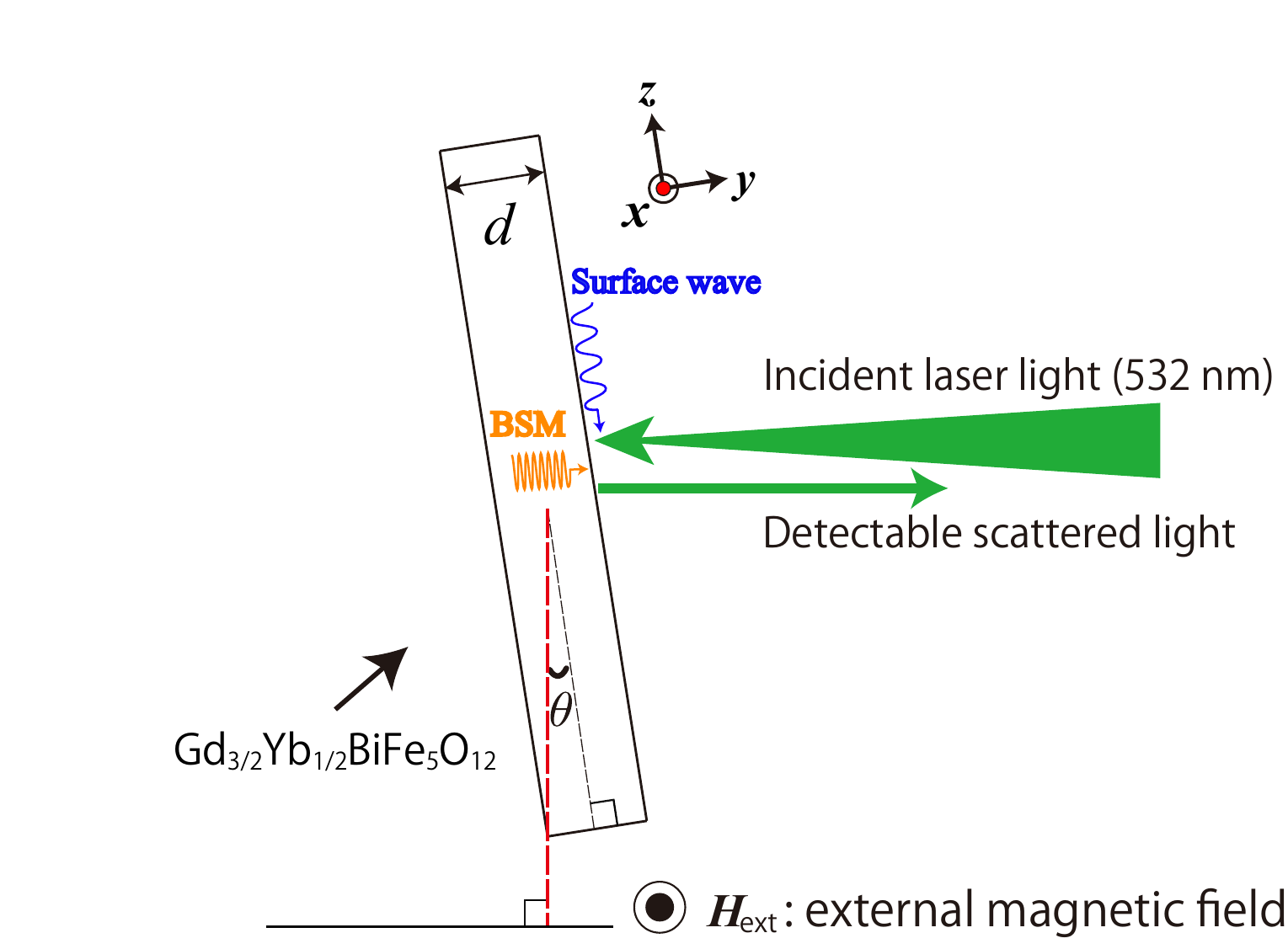}
		\caption{(Color online) The sample geometry. The reflected light is detected in a direction opposite to the incident light. $\theta$ ranges from 10$ ^\circ $ to 50$ ^\circ $.}
		\label{exp_geom}
	\end{figure}

	The BLS experiments were performed in the geometry shown in Fig. \ref{exp_geom}.
	The laser wavelength was 532 nm, the laser incidence angle $ \theta $ was varied between 10$ ^\circ $ and 50$ ^\circ $, and the power was $ \sim 100$ mW with a focus size of $ \sim 10\ \mu$m.
	Our setup was such that we could detect the backscattered light in a direction opposite to the incident light. An in-plane external magnetic field ($\mu_0H_{\text{ext}} >0.08 $ T) was applied along the $x$-direction, which is the Damon-Eshbach (DE) geometry. The field strength was large enough to saturate the in-plane magnetization. It should be noted that we did not apply any external excitation such as mirowaves or a current. However, thermally excited spin waves always exist at room temperature. These thermal spin waves scatter the incident light to be detected as BLS signals.
	
\begin{figure}[htbp]
		\includegraphics[width=10cm]{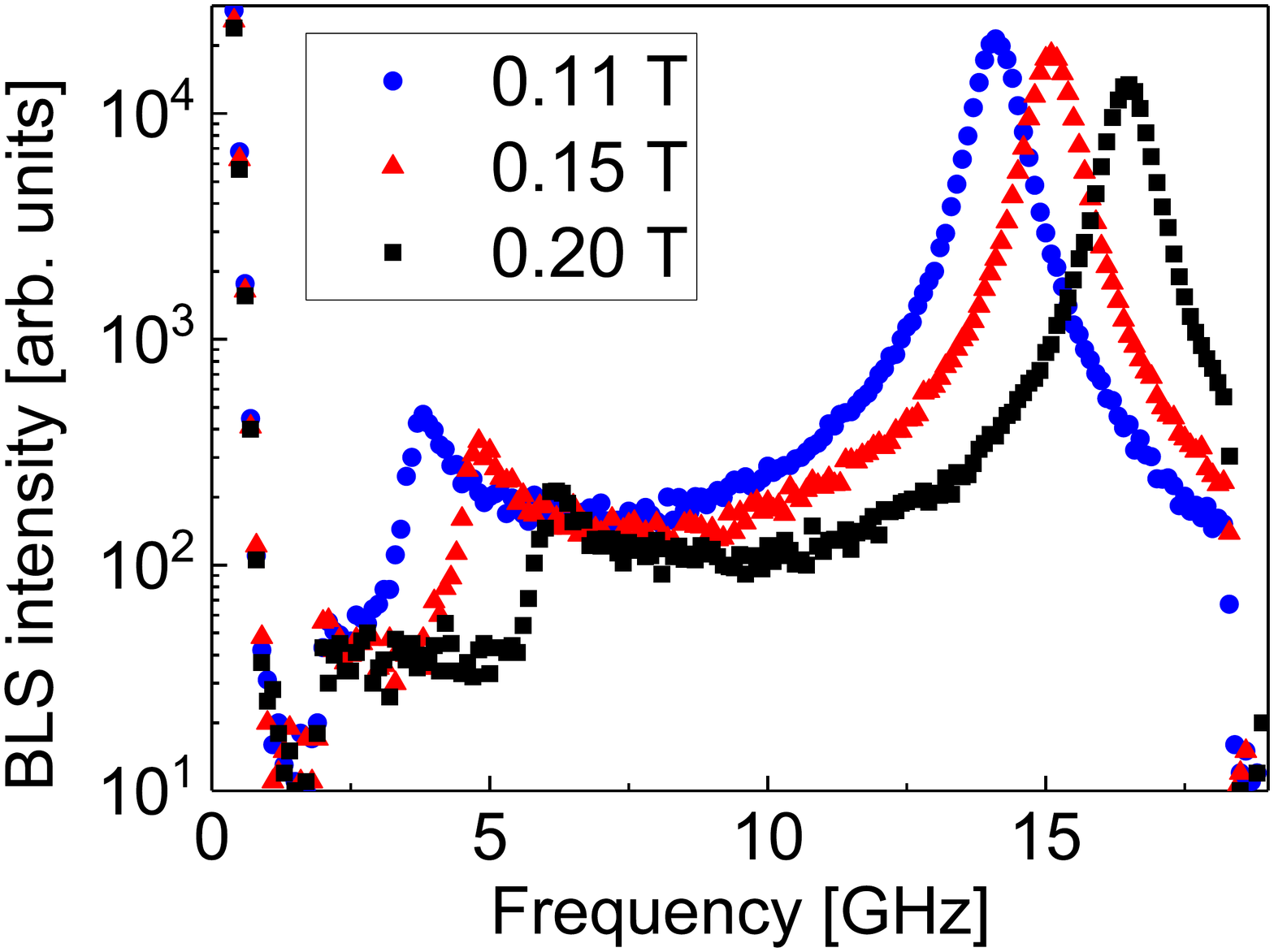}
		\caption{(Color online) BLS signals observed when $ \theta=10^\circ$ and $\mu_0H_{\text{ext}}=0.11,\ 0.15,\ 0.20$ T.}
		\label{bls_10deg_120mT}
	\end{figure}

	Figure \ref{bls_10deg_120mT} shows the BLS results at $ \theta=10^\circ$ and $\mu_0H_{\text{ext}}=0.11,\ 0.15,\ 0.20$ T. 
	We observed two different peaks: the first peak lies around 5 GHz and the second around 15 GHz. The intensity of the second peak was more than 50 times larger than the first peak. 
	The first peak is the surface spin wave mode in the DE geometry\cite{Eshbach60}.
	As shown later, the second peak is the BSM mode.
	
	From Eq. (3) in Ref. \cite{Lenk11} and Eq. (2) in Ref. \cite{SpinwaveConfinement}, one can obtain the dispersion of the dipole-exchange spin wave in the DE geometry as
	\begin{eqnarray}
	\frac{f^2}{\gamma^2\mu_0^2}=\left(H_{\text{ext}}+\frac{2A_{\text{ex}}}{\mu_0M_{\text{s}}}k^2\right)\left(H_{\text{ext}}+\frac{2A_{\text{ex}}}{\mu_0M_{\text{s}}}k^2+M_{\text{s}}-H_{\text{u}}\right)+\frac{M_{\text{s}}^2}{4}[1-\exp(-2kd)]\label{mssw_eqn}.
	\end{eqnarray}
	The wavenumber of the surface mode is $ k=k_{\text{surf}}=2\times k_{\text{I}}\sin\theta=2.362\times\sin\theta\times 10^7$ rad/m. This wavenumber is not negligible, compared to $ 2\pi/l_{\text{ex}}\sim 10^8$ rad/m, where the exchange length $ l_{\text{ex}}\equiv \sqrt{2A_{\text{ex}}/(\mu_0M_{\text{s}}^2)}$ is $\sim 30$ nm for a Bi-doped rare earth iron garnet\cite{Liu87}, so the observed surface mode is in the dipole-exchange regime. 
	
\begin{figure}[htbp]
		\includegraphics[width=10cm]{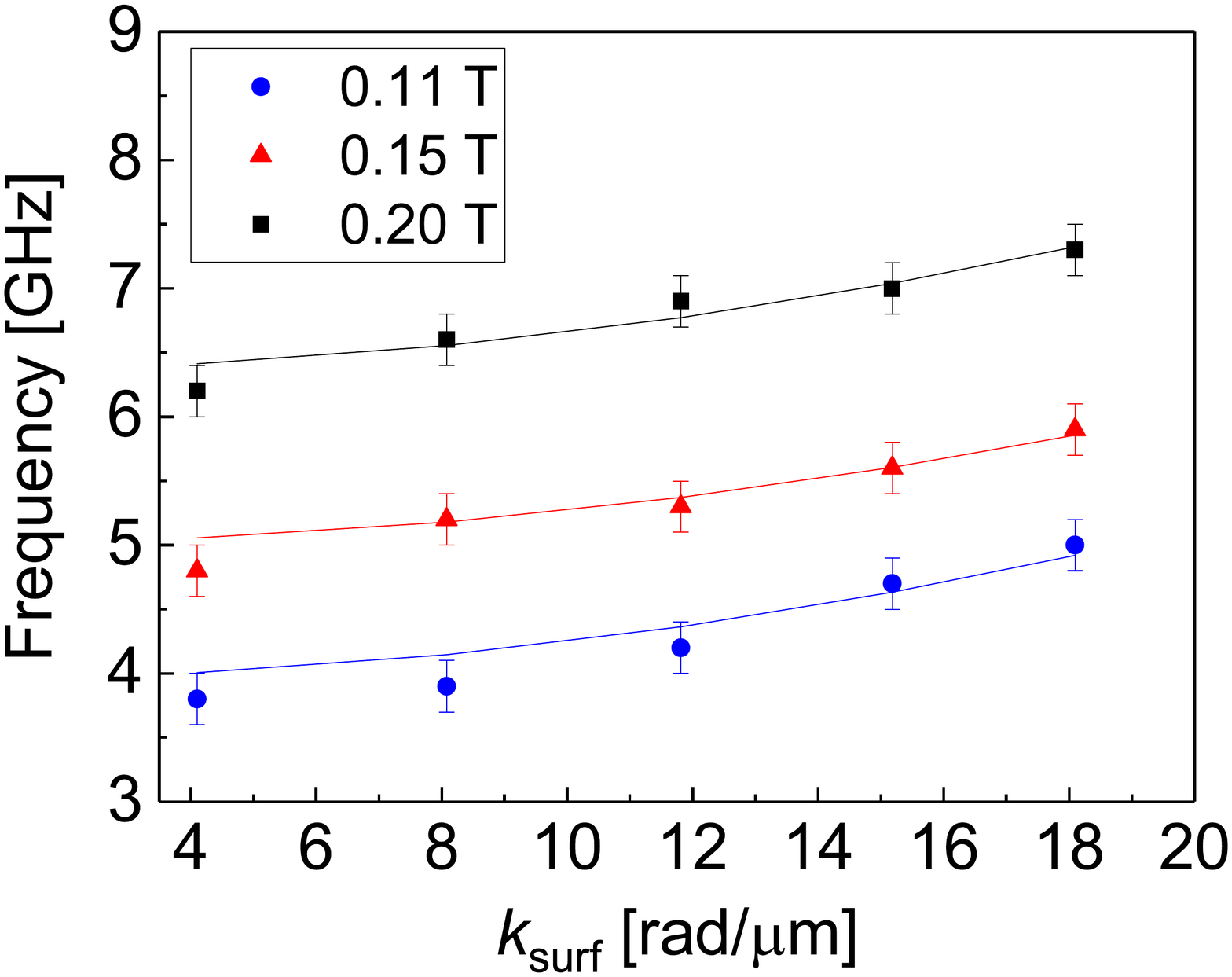}
		\caption{(Color online) BLS peak frequency (Dots) of the surface wave peak as a function of wavenumber. The fitted line (dashed line) was calculated using Eq. (\ref{mssw_eqn}).}
		\label{mssw_field_dep}
	\end{figure}

	Figure \ref{mssw_field_dep} shows the wavenumber dependence of the surface mode. 
	We fitted the frequency as a function of $k_{\text{surf}} $ using Eq. (\ref{mssw_eqn}). As a result, this surface mode was confirmed to be in the dipole-exchange regime, and we could estimate the exchange stiffness constant: $ A_{\text{ex}}=4.08\pm3.26,\ 3.78\pm3.03,\ 4.76\pm 3.12$ pJ/m for $ \mu_0H_{\text{ext}}=0.11,\ 0.15,\ 0.20$ T, respectively. The shown errors include the fitting variation within 99\% confidence errors and measurement errors. Thus, from the first peak, $ A_{\text{ex}}=4.32\pm3.57$ pJ/m.
	The large error shows the main limitation of this direct determination: The in-plane wave vector of the mode is very low and, consequently, the relative exchange contribution to the frequency is very small. As a result, the exchange stiffness constant cannot be determined with a high precision.
	
\begin{figure}[htbp]
		\includegraphics[width=10cm]{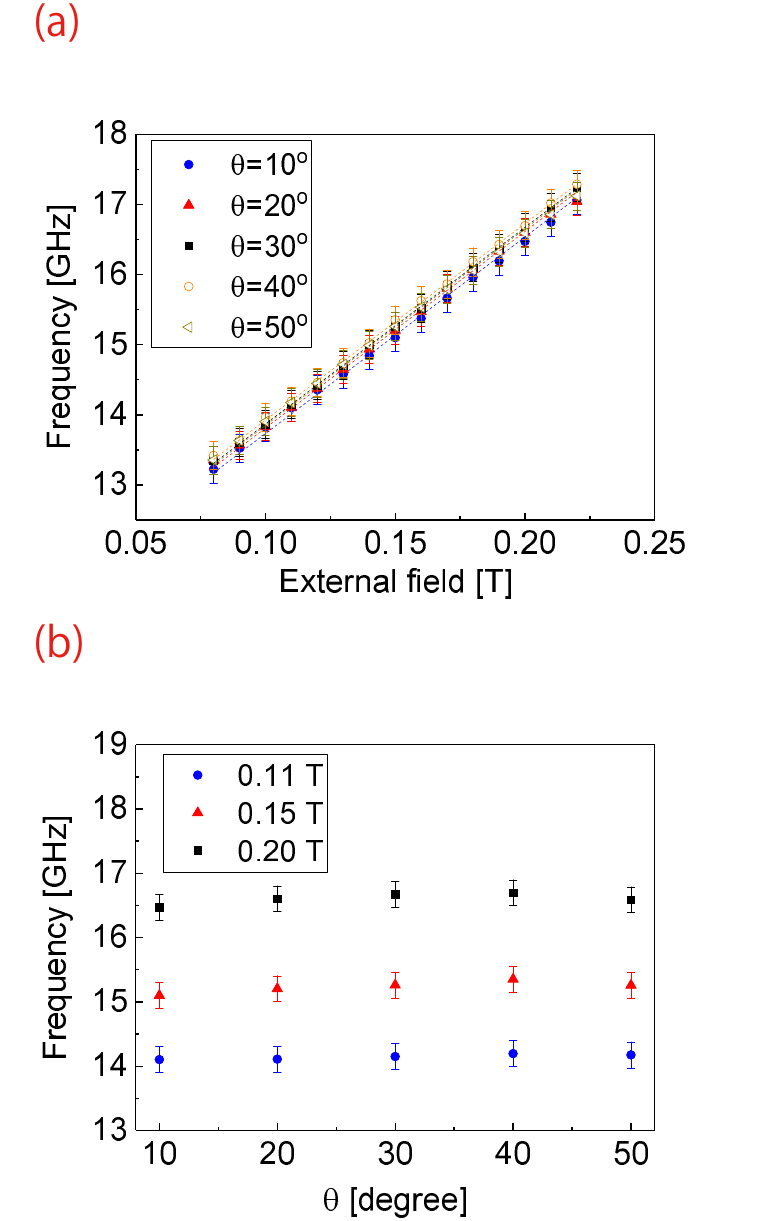}
		\caption{(Color online) (a) Peak frequency of the BSM peak (Dots) as a function of applied field for various tilting angles $ \theta $. Fitted results (dashed lines) using Eq. (\ref{mssw_eqn}). (b) BSM frequency as a function of incident angle when $\mu_0H_{\text{ext}}=0.11,\ 0.15,\ 0.20$ T.}
		\label{bsm_field_dep}
	\end{figure}

	Figure \ref{bsm_field_dep}(a) and (b) show the $H_{\text{ext}}  $ and $ \theta $ dependence of the second peak. The peak frequencies change with $H_{\text{ext}}  $, while they do not change with $ \theta $, which is expected for a BSM mode. The wavenumber of the BSM mode is $k_{\text{BSM}}=2nk_{\text{I}}=6.61\times 10^7$ rad/m, which indicates that the BSM mode is in the exchange-dominated regime.
	This wavenumber does not change with the incident angle since one can see that the variation of the frequencies in Fig. \ref{bsm_field_dep}(b) is within the experimental error. 
	
	As shown in Fig. \ref{exp_geom}, the direction of the observed BSM wavevector is perpendicular to the magnetization, and its dispersion is again expressed as Eq. (\ref{mssw_eqn}), where  $ \frac{M_{\text{s}}^2}{4} [1-\exp(-2k_{\text{BSM}}d)] $ is omitted because the BSM mode is a volume mode \cite{Stancil}.
	The fitting results, shown as dashed lines in Fig. \ref{bsm_field_dep}(a), confirm that the second peak is the exchange-dominated spin wave.
	The exchange stiffness constant is obtained as $A_{\text{ex}}=3.84\pm0.04,\ 3.87\pm0.04,\ 3.89\pm0.04,\ 3.91\pm0.04,\ 3.88\pm0.04$ pJ/m for $ \theta=10^\circ,\ 20^\circ,\ 30^\circ,\ 40^\circ,\ 50^\circ $, respectively.
	Thus, we obtained $ A_{\text{ex}}=3.87\pm0.08$ pJ/m, which is comparable with the result of YIG (3.7$ \pm $0.4 pJ/m\cite{Klingler15}). In addition, the accuracy of determination of $A_{\text{ex}}$ using the BSM mode is higher relative to that using the surface mode. This is because the BSM mode is the spin wave with the maximum wavenumber, i.e., the maximum exchange contribution that can be probed by BLS.
	
	Note that the present method to determine the exchange stiffness, $A_{\text{ex}}$, requires measuring the refractive index, and the accuracy of measurement directly affects the estimation of $A_{\text{ex}}$.
	Furthermore, this method utilizes magnetic BLS, so it is more suitable for materials with higher magneto-optical coupling, such as a Bi-doped rare earth iron garnet.
	
\begin{figure}[htbp]
		\includegraphics[width=10cm]{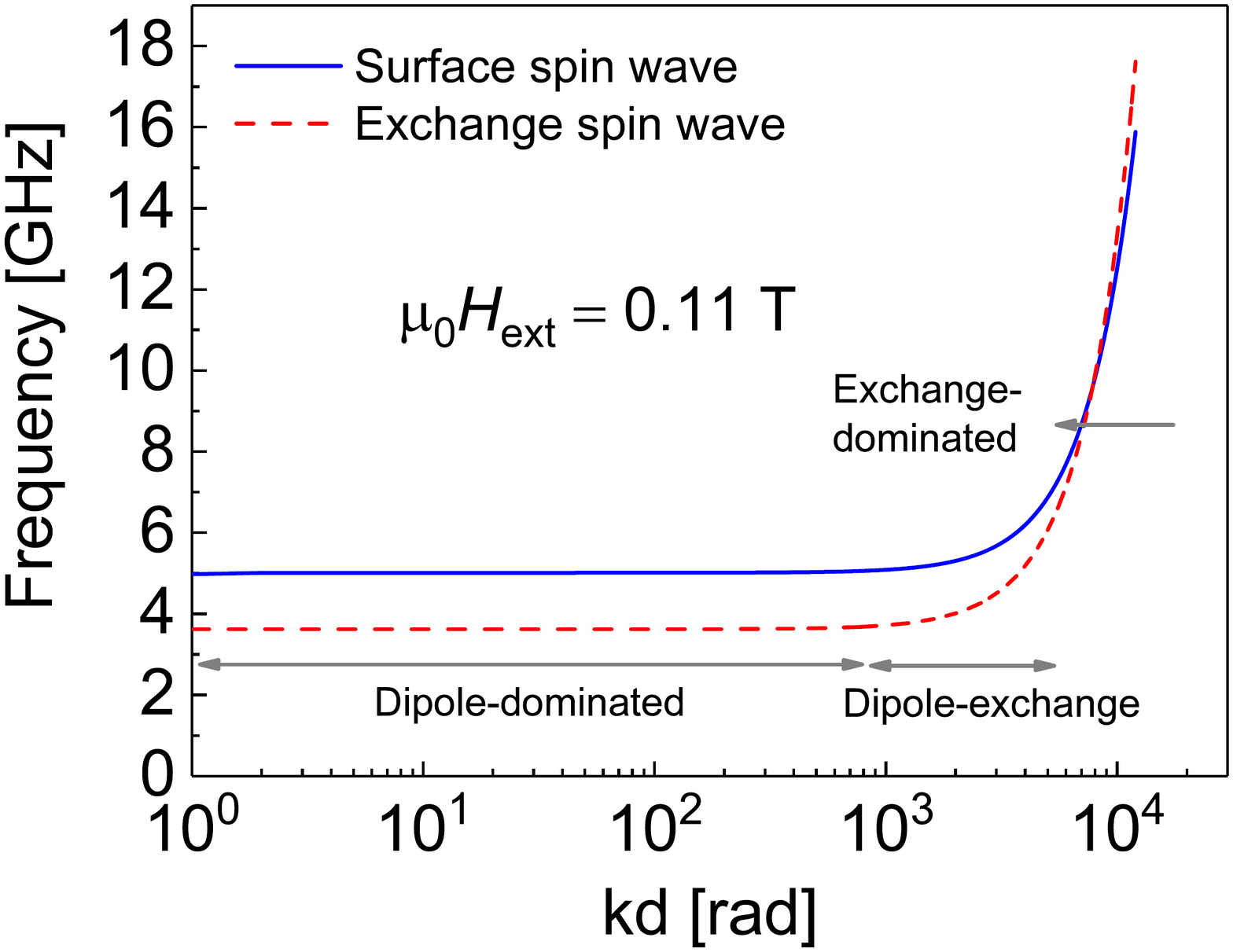}
		\caption{(Color online) The dispersion curve by using $ A_{\text{ex}} $. The blue line and red line show the dispersion curve of the surface mode and the exchange spin wave mode for $ \mu_0H_{\text{ext}}=0.11$ T, respectively.}
		\label{dispersion}
	\end{figure}

	Finally, using the obtained $ A_{\text{ex}}$, we plotted the dispersion curve of the surface mode and the exchange spin wave mode as shown in Fig. \ref{dispersion}. It was reconfirmed that the surface mode ($ k_{\text{surf}} d\sim3\times10^3$ rad) lies in the dipole-exchange regime, whereas the BSM mode ($ k_{\text{BSM}} d\sim9\times10^3$ rad) lies in the exchange-dominated regime.
	
	In conclusion, we have observed a BSM signal and a surface spin wave signal. The pronounced BSM signal enabled us to estimate the exchange stiffness constant $ A_{\text{ex}} $ as $3.87 \pm 0.08$ pJ/m, which is comparable to that of YIG. Since the BSM mode had a larger wavenumber than the surface mode and the PSSW mode, the exchange spin wave could be easily probed, which resulted in the accurate estimation of $A_{\text{ex}}$. This measurement technique of using the BSM mode can be applied for other optically transparent materials. We believe our result will make the determination of $ A_{\text{ex}}$ with high precision possible.
\section{Acknowledgement}
	KM acknowledges support from the JSPS Core-to-Core Program (A. Advanced Research
	Networks) during his stay at Technische Universit\"at Kaiserslautern. 
	Financial support by the Deutsche Forschungsgemeinschaft (DFG) via the Graduate School Materials Science in Mainz (MAINZ) through the Excellence Initiative (GSC 266) and the Project B01 of the SFB/TRR 173 Spin + X, the European Research Council Starting Grant 678309 MagnonCircuits, as well as by the Nachwuchsring of the TU Kaiserslautern is gratefully acknowledged.
	TS was financially supported by JSPS KAKENHI (numbers JP15H05454 and JP26103004).
	

\begin{thebibliography}{99}
		\bibitem{Stancil}
		D. D. Stancil and A. Prabhakar, 
		{\it Spin Waves: Theory and Applications}
		(Springer, Heidelberg, 2009) Chap. 5.
		
		\bibitem{SpinwaveConfinement}
		V. V. Kruglyak, P. S. Keatley, R. J. Hicken, J. R. Childress, and J. A. Katine,
		in {\it Spin Wave Confinement,} ed. S. O. Demokritov
		(Pan Stanford, 2008) 1st ed., Chap. 3.
		
		\bibitem{Bobeck1969}
		A. H. Bobeck, R. F. Fischer, A. J. Perneski, J. P. Remeika, and L. G. Van
		Uitert, IEEE Trans. Magn. {\bf 5}, 544 (1969).
		
		\bibitem{McMichael1997}
		R. D. McMichael and M. J. Donahue, IEEE Trans. Magn. {\bf 33}, 4167 (1997).
		
		\bibitem{Chappert2007}
		C. Chappert, A. Fert, and F. N. Van Dau, Nat. Mater. {\bf 6}, 813 (2007).
		
		\bibitem{Kittel1958}
		C. Kittel, Phys. Rev. {\bf 110}, 1295 (1958).
		
		\bibitem{Kalinikos1986}
		B. A. Kalinikos and A. N. Slavin, J. Phys. C: Solid State Phys. {\bf 19}, 7013 (1986).
		
		\bibitem{Kalinikos1990}
		B. A. Kalinikos, N. G. Kovshikov, P. A. Kolodin, and A. N. Slavin, Solid State Commun. {\bf 74}, 989 (1990).
				
		\bibitem{Vernon84}
		S. P. Vernon, S. M. Lindsay, and M. B. Stearns,
		Phys. Rev. B {\bf 29}, 4439 (1984).
		
		\bibitem{Liu96}
		X. Liu, M. M. Steiner, R. Sooryakumar, G. A. Prinz, R. F. C. Farrow, and G. Harp,
		Phys. Rev. B {\bf 53}, 12166 (1996).
		
		\bibitem{Raasch94}
		D. Raasch, J. Reck, C. Mathieu, and B. Hillebrands,
		J. Appl. Phys. {\bf 76}, 1145 (1994).
		
		\bibitem{Hamrle09}
		J. Hamrle, O. Gaier, S.-G. Min, B. Hillebrands, Y. Sakuraba, and Y. Ando,
		J. Phys. D: Appl. Phys. {\bf 42}, 084005 (2009).
		
		\bibitem{Gaier2009}
		O. Gaier, J. Hamrle, S. Trudel, B. Hillebrands, H. Schneider, and G. Jakob, J. Phys. D: Appl. Phys. {\bf 42}, 232001 (2009).
		
		\bibitem{Eyrich12}
		C. Eyrich, W. Huttema, M. Arora, E. Montoya, F. Rashidi, C. Burrowes, B. Kardasz, E. Girt, B. Heinrich, O. N. Mryasov, M. From, and O. Karis,
		J. Appl. Phys. {\bf 111}, 07C919 (2012).
		
		\bibitem{Klingler15}
		S. Klingler, A. V. Chumak, T. Mewes, B. Khodadadi, C. Mewes, C. Dubs, O. Surzhenko, B. Hillebrands, and A. Conca,
		J. Phys. D: Appl. Phys. {\bf 48}, 015001 (2015).
		
		\bibitem{Sebastian15}
		T. Sebastian, Y. Kawada, B. Obry, T. Br\"acher, P. Pirro, D. A. Bozhko, A. A. Serga, H. Naganuma, M. Oogane, Y. Ando, and B. Hillebrands,
		J. Phys. D: Appl. Phys. {\bf 48}, 164015 (2015).
		
		\bibitem{Shirane68}
		G. Shirane, V. J. Minkiewicz, and R. Nathans,
		J. Appl. Phys. {\bf 39}, 383 (1968).
		
		\bibitem{Passell1976}
		L. Passell, O. W. Dietrich, and J. Als-Nielsen, Phys. Rev. B {\bf 14}, 4897 (1976).
		
		\bibitem{Scholl04}
		A. Scholl, M. Liberati, E. Arenholz, H. Ohldag, and J. St\"ohr,
		Phys. Rev. Lett. {\bf 92}, 247201 (2004).
		
		\bibitem{Girt11}
		E. Girt, W. Huttema, O. N. Mryasov, E. Montoya, B. Kardasz, C. Eyrich, B. Heinrich, A. Yu. Dobin, and O. Karis,
		J. Appl. Phys. {\bf 109}, 07B765 (2011).
		
		\bibitem{Demokritov2001}
		S. O. Demokritov, B. Hillebrands, and A. N. Slavin, Phys. Rep. {\bf 348}, 441 (2001).
		
		\bibitem{Agrawal13}
		M. Agrawal, V. I. Vasyuchka, A. A. Serga, A. D. Karenowska, G. A. Melkov, and B. Hillebrands,
		Phys. Rev. Lett. {\bf 111}, 107204 (2013).
		
		\bibitem{Satoh12}
		T. Satoh, Y. Terui, R. Moriya, B. A. Ivanov, K. Ando, E. Saitoh, T. Shimura, and K. Kuroda,
		Nat. Photonics {\bf 6}, 662 (2012).
		
		\bibitem{Parchenko13}
		S. Parchenko, A. Stupakiewicz, I. Yoshimine, T. Satoh, and A. Maziewski,
		Appl. Phys. Lett. {\bf 103}, 172402 (2013).
		
		\bibitem{Yoshimine14}
		I. Yoshimine, T. Satoh, R. Iida, A. Stupakiewicz, A. Maziewski, and T. Shimura,
		J. Appl. Phys. {\bf 116}, 043907 (2014).
		
		\bibitem{Yoshimine17}
		I. Yoshimine, Y. Y. Tanaka, T. Shimura, and T. Satoh,
		EPL {\bf 117}, 67001 (2017).
		
		\bibitem{Chekhov18}
		A. L. Chekhov, A. I. Stognij, T. Satoh, T. V. Murzina, I. Razdolski, and A. Stupakiewicz, Nano Lett. {\bf 18}, 2970 (2018).
		
		\bibitem{Lenk11}
		B. Lenk, H. Ulrichs, F. Garbs, and M. M\"unzenberg,
		Phys. Rep. {\bf 507}, 107 (2011).
		
		\bibitem{Doormann84}
		V. Doormann, J.-P. Krumme, C.-P. Klages, and M. Erman,
		Appl. Phys. A {\bf 34}, 223 (1984).
		
		\bibitem{Eshbach60}
		J. R. Eshbach and R. W. Damon,
		Phys. Rev. {\bf 118}, 1208 (1960).
		
		\bibitem{Liu87}
		Y. Liu, D. Mou, X. Li, and P. Zhang,
		IEEE Trans. Magn. {\bf 23}, 3329 (1987).
		
	\end{thebibliography}
\end{document}